\documentclass{PoS}

\title{Phase structure of three flavor QCD in external magnetic fields using HISQ fermions}

\ShortTitle{QCD in external magnetic fields using HISQ}

\author{\speaker{Akio Tomiya} \\
        RIKEN/BNL Research center, Brookhaven National Laboratory, 
Upton, NY, 11973, USA\\
        Key Laboratory of Quark \& Lepton Physics (MOE) and Institute of Particle Physics, \\ Central China Normal University, Wuhan 430079, China\\
        E-mail: \email{akio.tomiya@riken.jp
}}
\author{Heng-Tong Ding, Xiao-Dan Wang, Yu Zhang\\
        Key Laboratory of Quark \& Lepton Physics (MOE) and Institute of Particle Physics, \\Central China Normal University, Wuhan 430079, China\\
        E-mail: \email{hengtong.ding@mail.ccnu.edu.cn}, \email{xiaodanwang@mails.ccnu.edu.cn}, \email{yuzhang@mails.ccnu.edu.cn} 
        }
\author{Swagato Mukherjee\\
        Department of Physics, Brookhaven National Laboratory, Upton, New York 11973-5000\\
        E-mail: \email{swagato@bnl.gov}
        }
\author{Christian Schmidt\\
        Fakult\"at f\"ur Physik, Universit\"at Bielefeld, D-33615 Bielefeld, Germany \\
        E-mail: \email{schmidt@physik.uni-bielefeld.de}
        }

\abstract{
We study the phase structure of QCD with three degenerate flavors in the external magnetic fields using highly improved staggered quarks (HISQ).
The simulations are performed on $16^3\times 6$ lattice.
In order to investigate the quark mass dependence of the chiral transition we choose the values of the bare quark masses 0.015 and 0.0009375 in the lattice unit, corresponding to $m_\pi=320$ MeV and 80 MeV in the continuum limit.
We found no indication of a first order phase transition in the current window of quark masses and external magnetic fields. 
Unlike to the case with standard staggered fermions inverse magnetic catalysis is always observed at about the critical temperature. 
The microscopic origin of this phenomena are further discussed by looking into the Dirac eigenvalue spectrum.
}

\FullConference{The 36th Annual International Symposium on Lattice Field Theory - LATTICE2018\\
		22-28 July, 2018\\
		Michigan State University, East Lansing, Michigan, USA.}
\RequirePackage{xspace} 
\usepackage{amsmath,amssymb}
\usepackage{mathrsfs}
\usepackage{bm}

\begin{document}

\newcommand{\mud}{m_\text{ud}}
\newcommand{\ms}{m_\text{s}}
\newcommand{\mq}{m_\text{q}}
%
\newcommand{\divergence}{\mathrm{div}\,}  
\newcommand{\grad}{\mathrm{grad}\,}  
\newcommand{\rot}{\mathrm{rot}\,}  
\newcommand{\tr}{\mathrm{tr}\,}  
\newcommand{\Tr}{\mathrm{Tr}\,}  
\newcommand{\Det}{\mathrm{Det}\,}  
\newcommand{\diag}{\mathrm{diag}\,}  
\newcommand{\re}{\mathrm{Re}\,} 
\newcommand{\sgn}{\mathrm{sgn}\,}   
\newcommand{\fsla}[1]{{\ooalign{\hfil/\hfil\crcr$#1$}}} 
\newcommand{\Ds}{\fsla{D}}  
\newcommand{\av}[1]{\left\langle #1 \right\rangle} 
\newcommand{\Bra}[1]{\left\langle #1 \right|} 
\newcommand{\Ket}[1]{\left| #1 \right\rangle} 
\renewcommand{\bar}[1]{\overline{#1}}%
\newcommand{\br}{\notag\\&\;\;\;\;\;\;}
\newcommand{\mass}{ am_\text{q} }
\newcommand{\fig}{Fig.\xspace}
\newcommand{\tab}{Tab.\xspace}
\newcommand{\ucite}[1]{\xspace\textcolor{red}{[#1]}}
\newcommand{\suthree}{SU(3)\xspace}
\newcommand{\uone}{U(1)\xspace}
%

%
%
%
\section{Introduction}
Mapping out QCD phase diagram is one of the most important goals in high energy nuclear physics.
QCD thermodaynamics has several controllable parameters, temperature $T$, baryon chemical potential 
and in addition quark masses $\mud,\; \ms$ for theoretical studies.
Quark mass is a relevant parameter and it affects the order of QCD phase transition
because chiral symmetry is controlled by the mass,
which is summarized in the Columbia plot~\cite{Pisarski:1983ms,Brown:1990ev}.
One can introduce an extra parameter, e.g. the background magnetic field $eB$ \cite{Costa:2017zey}, which can be generated in 
the early stage of relativistic heavy ion collision experiments.
The presence of the magnetic field breaks flavor symmetry and rotational symmetry, and the magnitude of the magnetic field is expected to reach to QCD scale so the chiral phase transition would be affected.

From lattice QCD simulations with external magnetic field at the physical pion mass, especially
using stout smeared staggered actions, the inverse magnetic catalysis
has been found around the pseudo critical temperature \cite{Bruckmann:2013oba}.
Recently, the origin of the inverse magnetic catalysis, in particular the relation between the decreasing of 
pseudo critical temperature and the non-monotonic behavior of chiral condensate around the pseudo critical 
temperature has been studied in details in $N_f$=2+1 QCD with a few values of heavy pions using  stout staggered fermions \cite{DElia:2018xwo}.
A previous lattice study with an effective model suggested a new critical end point along with the magnetic field. 
They employed stout smeared staggered quarks with physical pion mass for their lattice calculation
and gluonic effective model for predictions~\cite{Endrodi:2015oba}.
The new critical point is predicted to appear $eB \sim 10 \; \text{GeV}^2$ so it is practically difficult to reach such a strong magnetic field
using lattice calculations since the maximum of the magnetic field is bounded by the cutoff $a^{-1}$.
The cutoff at the critical point is $a^{-1} \sim 2.4$ GeV.
However,
there is a chance to observe the first oder phase transition
if a system is in proximity to a critical point.
This is because,
according to our previous work using the standard staggered fermions,
the magnetic field tends to make the phase transition stronger \cite{Tomiya:2017cey}.

In this work, we investigate mass degenerate three flavor QCD,
namely, $m = m_\text{ud}= m_\text{s}$ and around $SU(3)$ chiral limit.
We performed simulations with quark masses corresponding to pion masses of $320$ MeV and $80$ MeV in the continuum limit.
The results with the heavier quark is used to check the cutoff effects because that mass corresponds to similar mass regime
to the our previous work with the standard staggered fermion, while those with the
lighter quark are used to explore criticality as it is much closer to the first order regime compared to
the physical pion mass regime.

This proceedings is organized as follows.
In next section, we introduce our numerical setup,
which includes the implementation of the external field on the lattice 
with smeared links.
In section \ref{sec:spect}, we review a stochastic estimator
to calculate the Dirac spectrum with the external field.
In section \ref{sec:results}, we show our results.
In section \ref{sec:summary}, we summarize our observations.

\section{Setup} \label{sec:setup}
\subsection{Magnetic field with HISQ Dirac operator}
In this section we introduce our setup.
We employ Highly Improved Staggered Quarks (HISQ)
to suppress lattice artifacts including the taste violation effect.
HISQ is constructed by two of fat-7 smearing with re-unitarization between them.
The smeared links are obtained in the following way.
Firstly level one smeared links $V_\mu$ are constructed by fat-7 from thin \suthree links $U_\mu$.
Next, re-uniterized links $W_\mu$ are constructed by projecting $V_\mu$ on $U(3)$.
Finally, level two smeared links $X_\mu$ are constructed by fat-7 from thin \suthree links $W_\mu$ with
the Lepage term.
The HISQ Dirac operator are built by the Kogut-Suskind term with $X_\mu$ and
the Naik term with $W_\mu$.
Note that unlike the implementation of imaginary chemical potential,
the force term in HMC has to be modified because it depends on the coordinate.

The magnetic field only couples to quarks thus implementation is done just by replacing 
$X_\mu \to u_\mu X_\mu$ in the Kogut-Suskind term
and
$W_\mu \to u_\mu W_\mu$ in the Naik term.
Here $u_\mu$ represents \uone links with appropriate charge for each quark,
which is explained below.

On the lattice, the external \uone magnetic field is realized as a \uone link
and is constructed in the following way.
Finiteness of lattice size introduces an infrared cutoff to the \uone field \cite{al2009discrete}.
Let us denote the lattice size $(N_x,\; N_y,\; N_z,\; N_t)$ and coordinate as $n_\mu=0,\cdots, N_\mu-1$ ($\mu = x,\; y,\; z,\; t$).
The external magnetic field in $z$ direction $\vec{B}=(0,0,B)$ is described by the link variable $u_\mu(n)$ of the \uone field and 
$u_\mu(n)$ is expressed as follows,
\begin{align}
u_x(n_x,n_y,n_z,n_t)&=
\begin{cases}
\exp[-iq \hat{B} N_x n_y] \;\;&(n_x= N_x-1)\\
1 \;\;&(\text{otherwise})\\
\end{cases}\notag\\
u_y(n_x,n_y,n_z,n_t)&=\exp[iq \hat{B}  n_x],\label{eq:def_mag_u} \\
u_z(n_x,n_y,n_z,n_t)&=u_t(n_x,n_y,n_z,n_t)=1.\notag
\end{align}
Here $q$ is the electric charge of each quark
and $\hat{B} \equiv a^2 B$. 
%
One-valuedness of the one particle wave function along with a plaquette requires the Dirac quantization, 
\begin{align}
q\hat{B} = \frac{2\pi N_b}{N_x N_y} \label{eq:mag_to_Nb},
\end{align}
where $N_b \in { \boldsymbol Z}$ is the number of magnetic flux through unit area for $x$-$y$ plane.
The ultraviolet cutoff $a$ introduces also a periodicity of the magnetic field along with $N_b$.
Namely, a range
$0\leq N_b < {N_x N_y}/{4}$,
represents an independent magnitude of the magnetic field $B$.

\subsection{Lattice setup}
We employ the HISQ action with a tree-level Symanzik gauge action.
To generate configurations, we use the Rational Hybrid Monte-Carlo algorithm.
We perform multi-stream runs to increase the statistics.
The quark masses are degenerate but up type and down(strange) type quarks are assigned
different electromagnetic charges as those in the real world.
Our lattice size is $16^3 \times 6$ and $a^{-1} \sim 740$ MeV.
$\beta$ ranges are $[5.8,~ 6.2]$ and $[5.7,~ 6.05]$ for $m_\pi = 320 $ MeV
and $m_\pi = 80$ MeV, respectively.
We take $ 0 \leq N_b \leq 56$, which corresponds to $ 0 \leq \sqrt{eB} \lesssim 870$ MeV in the physical unit.

We measure standard observables, the chiral condensate and its susceptibility 
and the Binder cumulant \cite{binder1981critical} for the chiral condensate.
The Binder cumulant is defined as a function of $\beta$,
\newcommand{\deltapsi}{\delta\bar{\psi}\psi}
\begin{align}
B_4 (\beta) = \frac{ \av {(\deltapsi)^4} }{ \av{(\deltapsi)^2}^2 },
\end{align}
where $\deltapsi = \bar{\psi}\psi - \av{\bar{\psi}\psi} $.
The minimum of {Binder cumulant} $B_4$ indicates orders of phase transitions:
$B_4 =3$ corresponds to crossover,
$B_4 \sim 1.6$ {for the} second order phase transition with the Ising $\mathbb{Z}_2$ universality class,
$B_4 = 1$ {for the} first order phase transition \cite{Bazavov:2017xul}.

In addition, we measure three types of condensates, 
\begin{align}
\av{\bar{\psi} \psi}^\text{full,$f$}(B) &= \int \mathcal{D} U P[U;B] \; \Tr\left[ \frac{1}{ D^\text{$f$}[U;B] + m } \right], 
\label{eq:condensated_eB_dependence_full}
\\
\av{\bar{\psi} \psi}^\text{val,$f$}(B) &= \int \mathcal{D} U P[U;0] \; \Tr\left[ \frac{1}{ D^\text{$f$}[U;B] + m } \right], 
\label{eq:condensated_eB_dependence_sea}
\\
\av{\bar{\psi} \psi}^\text{sea,$f$}(B) &=  \int \mathcal{D} U P[U;B] \; \Tr\left[ \frac{1}{ D^\text{$f$}[U;0] + m } \right] ,
\label{eq:condensated_eB_dependence_val}
\end{align}
where
$
P[U;B] = \frac{1}{Z(B)} e^{-S_g[U]} \Det[D^\text{up}[U;B] + m]^{1/4} \Det[D^\text{down}[U;B] + m]^{1/2}
$
and $f$ denotes up and down quarks.
The flavor averaged chiral condensate is defined by
\begin{align}
\av{\bar{\psi} \psi}^{l}(B) = \left(\av{\bar{\psi} \psi}^{l,\text{up}}(B)  
+ \av{\bar{\psi} \psi}^{l,\text{down}}(B) 
\right)/2
\end{align}
for $l=\text{full, val, sea}$.
We calculate relative increases of the chiral condensate along with the magnetic field \cite{DElia:2011koc},
\begin{align}
r^l(B) = \frac{ \av{\bar{\psi} \psi}^{l}_\text{spec}(B)  - \av{\bar{\psi} \psi}^{l}_\text{spec}(0) }{ \av{\bar{\psi} \psi}^{l}_\text{spec}(0) }
\end{align}
where $\av{\bar{\psi} \psi}^{l}_\text{spec}(B)$ is defined using spectral summation with the Dirac spectrum
which is explained below.
They are related by $r^\text{full}(B) = r^\text{val}(B) + r^\text{sea}(B) + O(B^4)$ \cite{DElia:2011koc}.
This formula is only valid in $e\hat{B} \ll 1$ because it is based on the Taylor expansion with respect to $e\hat{B}$.
However, it can shed some light on the origin of the normal/inverse magnetic catalysis.

\section{Stochastic spectrum estimator}\label{sec:spect}
Here we review the stochastic spectrum estimator \cite{Giusti:2008vb,Fodor:2016hke, Cossu:2016eqs,deForcrand:2017cja}.
Let us denote $n[s,t]$ as the number of eigenvalues of a Hermitian operator $\tilde{D} = \tilde{D}(A)$ having eigenvalues in $[-1,1]$ in a range $[s,t]$.
We use a ramp function $h_{[s,t]}(x)$ with a support $[s,t]$, {\it i.e.}
\begin{align}
h_{[s,t]}(x) = 
\begin{cases}
1\;\;\;\; (s<x<t),\\
0\;\;\;\; \text{(otherwise)}.
\end{cases}
\end{align}
By definition, the number is
$n[s,t]
= 
\sum_j
\langle
h_{[s,t]}(\lambda_j^{\tilde{D}(A)})
\rangle_A$, where $\av{\cdots}_A$ is an ensemble average and
$\lambda_j^{\tilde{D}(A)}$ is $j$--th eigenvalue of  $\tilde{D}$ on a gauge configuration $A$.
Summation over all of eigenvalues
can be represented by trace operation and,
\begin{align}
n[s,t]
= 
\text{Tr}\Big[
\langle
h_{[s,t]}(\tilde{D})
\rangle_A
\Big] 
\approx
\frac{1}{N_r}\sum_{k=1}^{N_r}
\langle
\xi_k^\dagger h_{[s,t]}(\tilde{D}) \xi_k
\rangle_A
\approx \sum_{n=0}^{p} \gamma^{(n)}_{[s,t]}  
\frac{1}{N_r}\sum_{k=1}^{N_r}
\langle
\xi_k^\dagger T_n(\tilde{D}) \xi_k
\rangle_A, \label{eq:stochastic}
\end{align}
where $h_{[s,t]}(x) \approx \sum_n^{p} \gamma^{(n)}_{[s,t]} T_n(x) $ and $ T_n(x) $ is the Chebyshev polynomial.
$\xi_k$ is $k$--th random vector and $N_r$ is the number of random vectors.
During the derivation, we have approximated
trace operation as a Monte-Carlo average of random vectors.
In actual calculation we introduce the Jackson dumping factor
to suppress the Gibbs phenomena in the most
right hand side in \eqref{eq:stochastic} as in Ref.\cite{Cossu:2016eqs}.

In order to apply the Chebyshev expansion to calculate
the Dirac spectrum, we take
\begin{align}
\tilde{D} = \frac{D^\dagger D - (\lambda^{D^\dagger D}_\text{max}+\lambda^{D^\dagger D}_\text{min})/2 }{
(\lambda^{D^\dagger D}_\text{max}-\lambda^{D^\dagger D}_\text{min})/2
},
\end{align}
where $\lambda^{D^\dagger D}_\text{max}$ and $\lambda^{D^\dagger D}_\text{min}$ are the maximum and minimum
eigenvalues of a positive Hermitian operator $D^\dagger D$, respectively.
Thus the range of eigenvalue of $\tilde{D}$ is restricted in 
$[-1, 1]$.
By taking the square root of ${\lambda^{D^\dagger D}}$,
we reconstruct the Dirac spectrum through the following equation,
\newcommand{\lamdd}{\lambda^{D^\dagger D}}
\begin{align}
\rho(\lambda) 
\equiv \rho(\sqrt{\lamdd}, \delta) 
= \frac{1}{2V}\frac{n[ s,t] }{\delta},
\end{align}
where  $\sqrt{\lambda^{D^\dagger D}} = \sqrt {(\frac{\lambda^{D^\dagger D}_{max}- \lambda^{D^\dagger D}_{min}}{2})s + \frac{\lambda^{D^\dagger D}_{max} + \lambda^{D^\dagger D}_{min}}{2} }$ and $ \sqrt{\lambda^{D^\dagger D}} + \delta = \sqrt {(\frac{\lambda^{D^\dagger D}_{max} - \lambda^{D^\dagger D}_{min}}{2})t + \frac{\lambda^{D^\dagger D}_{max} + \lambda^{D^\dagger D}_{min}}{2} }$. Here $V$ is the volume of the system and for the step size $\delta$ we have $\delta = 0.0005$.
The eigenvalues are $D \psi_j = i \lambda_j \psi_j$ and 
\begin{align}
\rho(\lambda) = \frac{1}{V}
\sum_j \av{\delta(\lambda-\lambda_j^A) }_A,
\end{align}
where $\psi_j $ is an eigenvector.
Using the Dirac spectrum, one can reconstruct the chiral condensate,
\begin{align}
\av{\bar{\psi}\psi}_\text{spec} = \int_0^\infty d\lambda \frac{2m }{\lambda^2 + m^2} \rho(\lambda), 
\end{align}
except for finite volume corrections.
This $\av{\bar{\psi}\psi}_\text{spec}$ is used in the calculation of the relative increase of chiral condensate $r$.

\section{Results} \label{sec:results}
Here we show our results for $m_\pi = 320$ MeV.
In general $m_\pi = 80$ MeV system shows qualitatively similar results with $m_\pi = 320$ MeV system except for behavior of the pseudo critical temperature in external magnetic fields so we just address some results on $m_\pi = 80$ MeV system later.

\begin{figure}[htbp]
        \begin{center}
              \begin{minipage}{0.49\hsize }
                \begin{center}
                        \includegraphics[width=\hsize, bb=0 0 270 141]{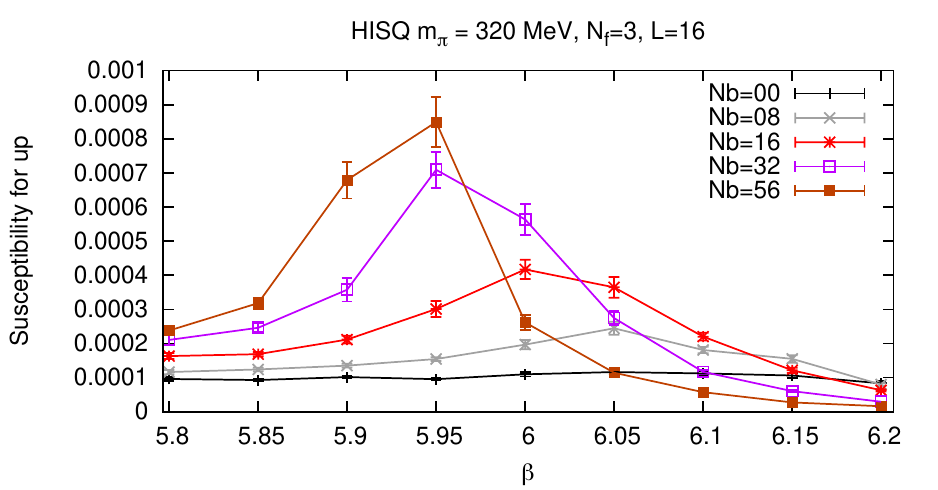}
                \end{center}
              \end{minipage}
              \begin{minipage}{0.49\hsize }
                \begin{center}
                        \includegraphics[width=\hsize, bb=0 0 270 141]{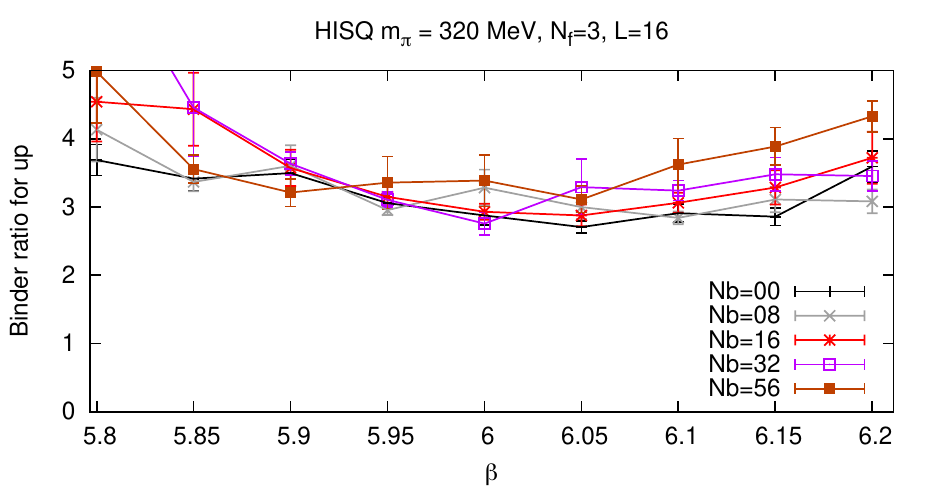}
                \end{center}
              \end{minipage}
        \end{center}
\caption{
Results for basics observables with $m_\pi = 320$ MeV.
Left panel shows the chiral susceptibility. For $N_b\neq 0$, we observe small peaks while 
it does not appear for $N_b = 0$.
Right panel shows the Binder cumulant for the chiral condensate.
There are no signal for chiral phase transition.
\label{fig:result_basic_mpi320_HISQ} }
\end{figure}

Our results for the chiral condensate susceptibility (left panel) and corresponding Binder cumulant (right panel) for $m_\pi = 320$ MeV are shown in \fig \ref{fig:result_basic_mpi320_HISQ}. It can be clearly seen that the peak location of the chiral susceptibility moves to lower temperature in stronger magnetic field and the peak heights increases with the magnetic field. The Binder cumulant shows that the system lies in the crossover regime. 

In Fig. 2 we show the Dirac spectra with $m_\pi = 320$ MeV at just above the critical temperature.
Dirac spectra with the magnetic field in both sea and dynamical quarks, sea quark only and valence quark only, obtained using \eqref{eq:condensated_eB_dependence_full}, \eqref{eq:condensated_eB_dependence_sea} and
\eqref{eq:condensated_eB_dependence_val}, are shown from left to right panel in Fig. 2.
For $N_b>0$, left four panels are consistent with inverse magnetic catalysis
while valence ones only show normal magnetic catalysis behavior. 

The relative increase of chiral condensate $r$ with $m_\pi = 320$ MeV in two different phases is shown in Fig. 3. The left panel is for the low temperature phase. All of them  monotonically increases as a function of $N_b$, namely they show normal magnetic catalysis. Right panel shows the $r$ dependence on $N_b$ in the high temperature phase. We can see that contributions from the dynamic sea quarks show non-montonical behavior. This indicates that the inverse magnetic catalysis seems to originate from contribution from dynamic sea quarks. This is consistent with the results from Ref.~\cite{Bruckmann:2013oba}. 

\begin{figure}[htbp]
        \begin{center}
                \begin{center}
                        \includegraphics[width=\hsize, bb=0 0 1000 550]{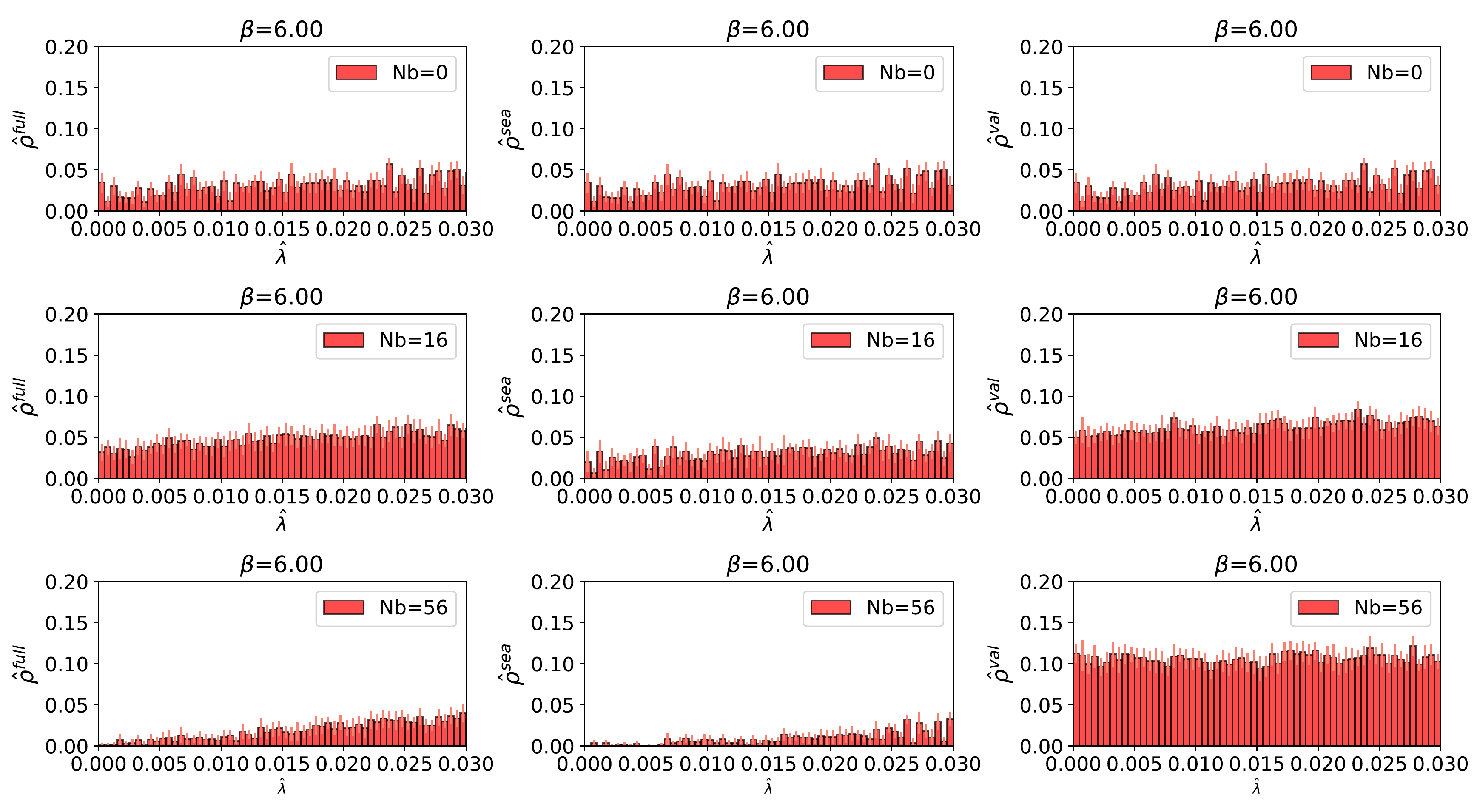}
                \end{center}
        \end{center}
\caption{
Dirac spectrum with $m_\pi = 320$ MeV at just above the critical temperature. 
From left to right, Dirac spectrum with the magnetic field both in sea and dynamical quarks, sea quark and valence quark (see text),
respectively. From top to bottom, $N_b=0, 16, 56$. Top plots are identical but are shown just for comparison.
For $N_b>0$, left four panels are consistent with inverse magnetic catalysis
while valence ones only show normal magnetic catalysis behavior.
\label{fig:result_mpi320_HISQ} }
\end{figure}

\begin{figure}[htbp]
        \begin{center}
              \begin{minipage}{0.45\hsize }
                \begin{center}
                        \includegraphics[width=\hsize, bb=0 0 504 504]{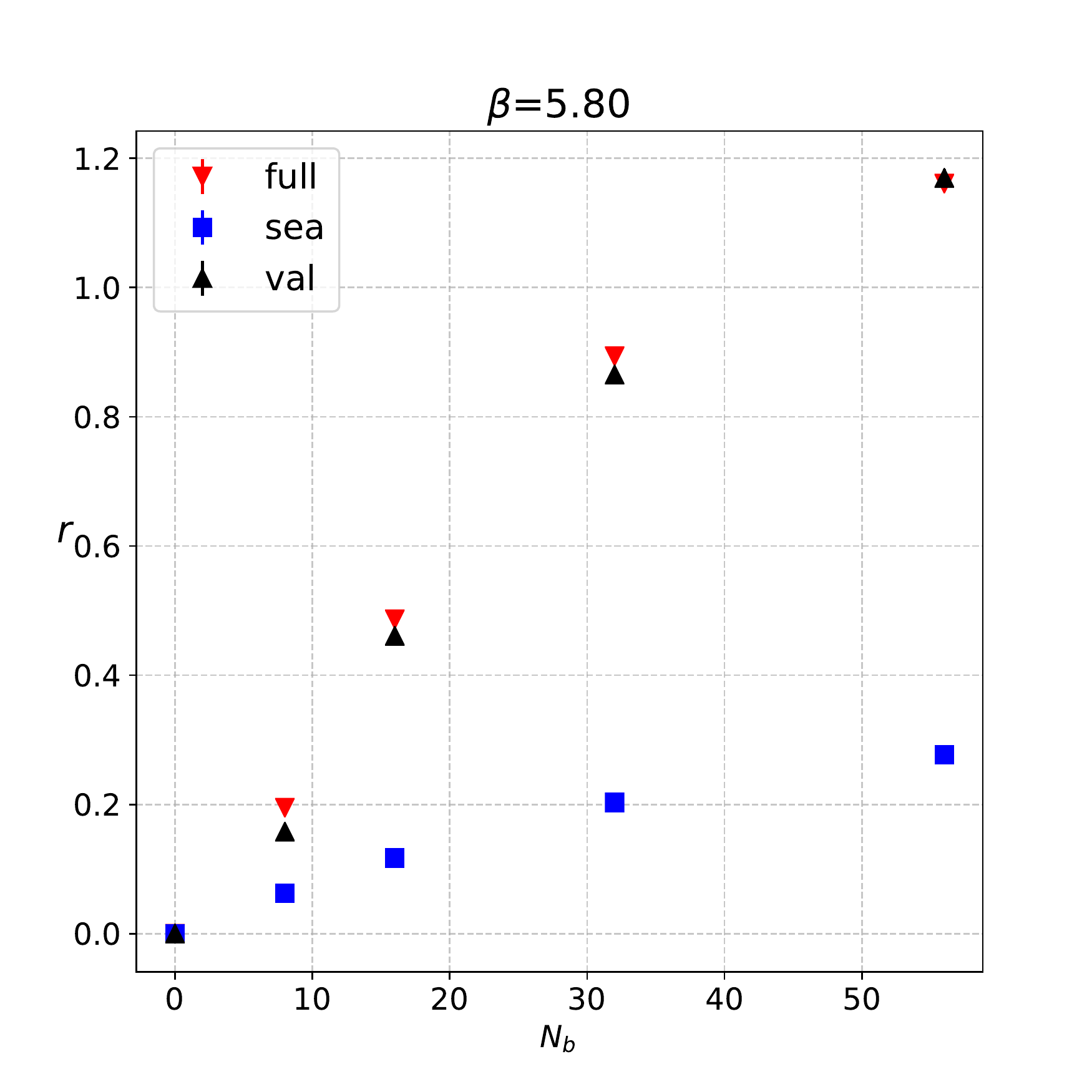}
                \end{center}
              \end{minipage}
              \begin{minipage}{0.45\hsize }
                \begin{center}
                        \includegraphics[width=\hsize, bb=0 0 504 504]{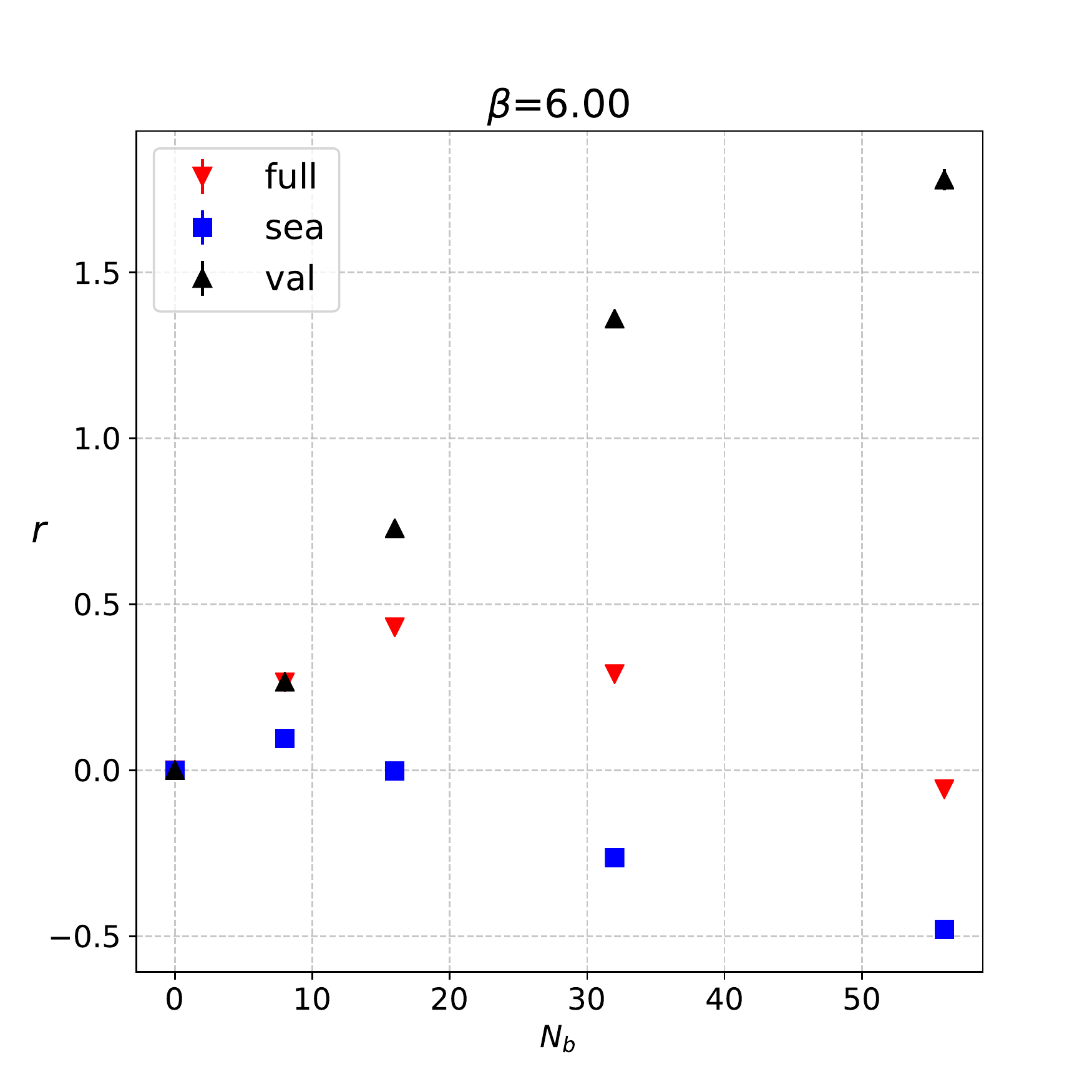}
                \end{center}
              \end{minipage}
        \end{center}
\caption{
The relative increase of chiral condensate $r$ with $m_\pi = 320$ MeV in two different phases.
Left panel is for low temperature phase. All of them are monotonically increased as a function of $N_b$,
namely they show normal magnetic catalysis.
Right panel shows $r$ dependence on $N_b$ in the high temperature phase.
Except for $r$ which contain magnetic field effect only in the probe,
they show inverse magnetic catalysis.
\label{fig:result_mpi320_HISQ_r} }
\end{figure}

We also measure the chiral condensate and its susceptibility for
$m_\pi = 80$ MeV system.
The susceptibility shows nontrivial dependence of peak location on 
$N_b$, namely, first the critical temperature increases and decreases.
Except for nontrivial dependence of the susceptibility for the
magnetic field, results shows qualitatively similar behavior.
It might be physical phenomena around chiral limit but
$m_\pi L  \sim 1.7 < 4$, so it could be finite volume effects and
further investigations are needed.
%
We observe that the Dirac spectrum is qualitatively similar to the heavier case.

\section{Summary} \label{sec:summary}
We have performed simulations of $N_f=3$ QCD in external magnetic fields using the HISQ action on $16^3\times6$ lattices. Two different values of quark masses are chosen corresponding to $m_\pi$=320 MeV and 80 MeV. We have observed inverse magnetic catalyses in both cases of $m_\pi = 320$ MeV and $80$ MeV.
Except for the dependence of the pseudo critical temperature on the background magnetic field,
two systems show qualitatively similar behavior.
On the other hand, $m_\pi = 80$ MeV data shows non monotonic dependence on the external field. However, this result is suspected coming from finite volume effects and
further investigation is needed.

\section*{Acknowledgement}
We would like to thank a lattice group in E\"otv\"os university for fruitful discussions.
The work of AT was supported in part by NSFC under grant no. 11535012
and the RIKEN Special Postdoctoral Researcher program. The numerical simulations have been performed on the GPU cluster in the Nuclear Science Computing Center at CCNU, and Tianhe II supercomputing center in Guangzhou.
This work was supported through Contract No. de-sc0012704 with the U.S. Department of Energy, and through the Scientific Discovery through Advanced Computing (SciDAC) program funded by the U.S. Department of  Energy, Office of Science, Office of Nuclear Physics.
CS acknowledge support by the Deutsche Forschungsgemeinschaft (DFG, German Research Foundation) -- project number 315477589 -- TRR 211.

\bibliographystyle{ieeetr}
\bibliography{ref}

\begin{thebibliography}{10}

\bibitem{Pisarski:1983ms}
Robert~D. Pisarski and Frank Wilczek.
\newblock {Remarks on the Chiral Phase Transition in Chromodynamics}.
\newblock {\em Phys. Rev.}, D29:338--341, 1984.

\bibitem{Brown:1990ev}
Frank~R. Brown, Frank~P. Butler, Hong Chen, Norman~H. Christ, Zhi-hua Dong,
  Wendy Schaffer, Leo~I. Unger, and Alessandro Vaccarino.
\newblock {On the existence of a phase transition for QCD with three light
  quarks}.
\newblock {\em Phys. Rev. Lett.}, 65:2491--2494, 1990.

\bibitem{Costa:2017zey}
Pedro Costa, Marcio Ferreira, and Constanca Providencia.
\newblock {Magnetized QCD phase diagram: critical end points for the strange
  quark phase transition driven by external magnetic fields}.
\newblock {\em PoS}, Hadron2017:161, 2018.

\bibitem{Bruckmann:2013oba}
Falk Bruckmann, Gergely Endrodi, and Tamas~G. Kovacs.
\newblock {Inverse magnetic catalysis and the Polyakov loop}.
\newblock {\em JHEP}, 04:112, 2013.

\bibitem{DElia:2018xwo}
Massimo D'Elia, Floriano Manigrasso, Francesco Negro, and Francesco Sanfilippo.
\newblock {QCD phase diagram in a magnetic background for different values of
  the pion mass}.
\newblock {\em Phys. Rev.}, D98(5):054509, 2018.

\bibitem{Endrodi:2015oba}
Gergely Endrodi.
\newblock {Critical point in the QCD phase diagram for extremely strong
  background magnetic fields}.
\newblock {\em JHEP}, 07:173, 2015.

\bibitem{Tomiya:2017cey}
Akio Tomiya, Heng-Tong Ding, Swagato Mukherjee, Christian Schmidt, and Xiao-Dan
  Wang.
\newblock {Chiral phase transition of three flavor QCD with nonzero magnetic
  field using standard staggered fermions}.
\newblock {\em EPJ Web Conf.}, 175:07041, 2018.

\bibitem{al2009discrete}
MH~Al-Hashimi and U-J Wiese.
\newblock Discrete accidental symmetry for a particle in a constant magnetic
  field on a torus.
\newblock {\em Annals of physics}, 324(2):343--360, 2009.

\bibitem{binder1981critical}
Kurt Binder.
\newblock Critical properties from monte carlo coarse graining and
  renormalization.
\newblock {\em Physical Review Letters}, 47(9):693, 1981.

\bibitem{Bazavov:2017xul}
A.~Bazavov, H.~T. Ding, P.~Hegde, F.~Karsch, E.~Laermann, Swagato Mukherjee,
  P.~Petreczky, and C.~Schmidt.
\newblock {Chiral phase structure of three flavor QCD at vanishing baryon
  number density}.
\newblock {\em Phys. Rev.}, D95(7):074505, 2017.

\bibitem{DElia:2011koc}
Massimo D'Elia and Francesco Negro.
\newblock {Chiral Properties of Strong Interactions in a Magnetic Background}.
\newblock {\em Phys. Rev.}, D83:114028, 2011.

\bibitem{Giusti:2008vb}
Leonardo Giusti and Martin Luscher.
\newblock {Chiral symmetry breaking and the Banks-Casher relation in lattice
  QCD with Wilson quarks}.
\newblock {\em JHEP}, 03:013, 2009.

\bibitem{Fodor:2016hke}
Zoltan Fodor, Kieran Holland, Julius Kuti, Santanu Mondal, Daniel Nogradi, and
  Chik~Him Wong.
\newblock {New approach to the Dirac spectral density in lattice gauge theory
  applications}.
\newblock {\em PoS}, LATTICE2015:310, 2016.

\bibitem{Cossu:2016eqs}
Guido Cossu, Hidenori Fukaya, Shoji Hashimoto, Takashi Kaneko, and Jun-Ichi
  Noaki.
\newblock {Stochastic calculation of the Dirac spectrum on the lattice and a
  determination of chiral condensate in 2+1-flavor QCD}.
\newblock {\em PTEP}, 2016(9):093B06, 2016.

\bibitem{deForcrand:2017cja}
Philippe de~Forcrand and Benjamin Jager.
\newblock {Alternatives to the stochastic ``noise vector'' approach}.
\newblock {\em EPJ Web Conf.}, 175:14022, 2018.

\end{thebibliography}

\end{document}